**Visualizing Electrical Breakdown and ON/OFF States in Electrically Switchable Suspended Graphene Break Junctions**

Hang Zhang,[1,†] Wenzhong Bao,[1,†] Zeng Zhao,[1] Jhao-Wun Huang,[1] Brian Standley,[2] Gang Liu,[1] Fenglin Wang,[1] Philip Kratz,[1] Lei Jing,[1] Marc Bockrath,[1,2,*] and Chun Ning Lau[1,*]

[1]Dept. of Physics and Astronomy, University of California, Riverside, CA, 92521,USA
[2]Dept. of Applied Physics, California Institute of Technology, Pasadena, California 91125,USA
[†] These authors contribute equally to this work.

**Abstract**

Narrow gaps are formed in suspended single to few layer graphene devices using a pulsed electrical breakdown technique. The conductance of the resulting devices can be programmed by the application of voltage pulses, with a voltage of 2.5V~4.5V corresponding to an ON pulse and voltages ~8V corresponding to OFF pulses. Electron microscope imaging of the devices shows that the graphene sheets typically remain suspended and that the device conductance tends to zero when the observed gap is large. The switching rate is strongly temperature dependent, which rules out a purely electromechanical switching mechanism. This observed switching in suspended graphene devices strongly suggests a switching mechanism via atomic movement and/or chemical rearrangement, and underscores the potential of all-carbon devices for integration with graphene electronics.

Long term archival information storage is an open challenge with a variety of approaches proposed,[1] including periodic migration to new media. However, such approaches require constant effort to avoid loss due to data errors caused by finite storage medium lifetime. A nonvolatile means of storing information densely which is stable for extended time periods is therefore highly desirable. One approach to address this issue is to store information in the arrangement of atoms rather than electrical charges. Previous work has shown the electromigration of metallic particles in multi-walled carbon nanotubes (MWNTs).[2,3] More recently, conductance switching was observed in graphene and graphitic break junctions, and the proposed switching mechanism was the formation and breakdown of carbon chains or filaments.[4,5] However, later works reported cyclable conductance switching in devices solely consists of electrodes on $SiO_2$ interrupted by nanogaps,[6,7] thus it remains controversial whether the swiching behavior in substrate-supported devices is intrinsic or merely arising from the underlying substrate.

To determine the role of the substrate in producing the switching behavior, and to gain more insight into electromigration and switching in graphene, we study switching in suspended graphene layers that are isolated from the substrate.[8-10] We demonstrate switching in such suspended graphene break junctions has similar behavior to that of substrate-supported devices. The switches are formed by breakdown of graphene to leave a small gap, using a sequential breakdown technique. The junction resistance can be controlled by the application of voltage pulses, with 4V corresponding to an ON pulse that decreases the device resistance and 8V corresponding to an OFF pulse that increases the device resistance. The devices can be cycled up to hundreds of times before becoming inoperative. SEM imaging of the gap in the ON and OFF states show a larger gap in the OFF state than the ON state. The similarity of the switching characteristics in suspended devices strongly suggests that the switching mechanism is the same as on substrate supported samples (though oxide breakdown cannot be unequivocally excluded). The switching rate depends strongly on temperature, which indicates that atomic motion, chemical rearrangement, or both are essential to the switching mechanism.

---

[*] Email: marc.bock@ucr.edu, lau@physics.ucr.edu

Few-layer graphene sheets (typically 1-5 layers thick) are mechanically exfoliated over Si substrates covered by a layer of 300nm-thick of $SiO_2$ and identified by optical microscopy.[11] The switch devices are fabricated with two different methods: (I) Using electron beam lithography (EBL) to fabricate the device on a graphene sample on the $SiO_2$/Si substrate, and then etching ~120nm $SiO_2$ under the graphene flakes by dipping the device into a buffered oxide etch solution.[9,10] (II) Graphene is exfoliated onto substrates that are pre-patterned with trenches, which are 250nm deep and 2.5-5 μm wide. Source and drain electrodes that consist of 10 nm of Ti and 70 nm of Au are deposited by evaporation through a shadow mask,[12] which is aligned carefully with the edge of the trenches. This approach minimizes the risk of sample collapse during the conventional EBL procedure. A completed device is shown in the inset to Fig. 1a, while the main panel plots the current *I* and the voltage *V* applied to the device versus time. *I* is approximately linear in *V* up to ~0.5V, with a two –terminal conductance *G* ~ 0.5 mS.

To create the switch, we perform the breakdown step using a modification of the electromigration technique that was reported previously.[4] The completed device is placed in a high vacuum, typically ~$10^{-6}$ Torr, and voltage pulses (~4 V, typical duration 0.1 s) are applied. These pulses sequentially reduce the device conductance until it becomes zero (Fig. 1b). The typical breaking current density is ~2 mA/μm. Figure 2a-c show the breakdown of another device. The device began its breakdown near the geometric center, and the broken region expanded upon further application of voltage until the breakdown was complete (Fig. 2c). An in situ SEM video of the breakdown process is available as supporting information. While breakdown often began at the center, for some samples, the breakdown began at the edge, but still centrally located relative to the electrodes. This suggests that the device temperature plays a critical role in the breakdown,[13,14] though it is not the only factor. Other factors, such as fluctuations and defects are also likely to be important parameters during electromigration.

Fig. 3a shows the *IV* characteristic of a device following a successful breakdown. When the voltage is swept from -8 V to +8V, at first the current is zero. When the voltage is ~4 V, the current rises, often by a series of step-like current jumps, indicating that the device is entering the "on" state. As the voltage is raised further, the device conductance decreases to zero, turning the device "off." The sweep in the reverse direction from +8 V to -8 V shows similar behavior. We note that the high vacuum is crucial for producing an operable switch device, presumably because of the removal of molecules that may contaminate or react with graphene devices, such as oxygen or water vapor. To exclude the possibility that the substrate contributes to the measured current by a parallel conduction path, we made a control device with the same exact fabrication procedure, including electrodes that are partially released from the substrates via etching, but without the graphene. These devices showed little or no measurable current up to the maximum voltage (10 V) used in our experiments (Fig. 3b). Breakdown did occur, however, at significantly higher voltages ~210 V. This demonstrates the presence of the graphene layer is essential for obtaining a measurable current within the 10V voltage range employed in the experiment.

The switching behavior in the suspended devices was reproducible, and the most stable devices could be switched many times. The switching behavior is shown in fig. 3c. Applying an "ON" pulse of 4V switches the device "ON", with typical is conductance $G$~25-40 µS; an "OFF" pulse of 8V, on the other hand, reduces the conductance to ~1 µS. This behavior in suspended devices is very similar to that from single-layer graphene devices that are supported on the $SiO_2$ substrates.[4] This strongly suggests that the switching mechanisms are the same in both cases, and that switching occurs in an all-carbon device. Notably, these switches are non-volatile and robust. For suspended devices, they can be switched through up to hundreds of cycles. As a demonstration, the device behavior after 300 cycles is shown in the right panels of fig. 3c, which is very similar to that of the first two cycles. However it was found that the number of switching cycles before device failure was less than that of those on substrates, as one may expect for the greater degree of fragility of suspended devices.

To understand better how the switching behavior occurred, we also imaged a different device by scanning electron microscopy (SEM) in both the ON and OFF states (Fig. 3d). We note that imaging the device *during* the switching cycles results in device degradation and failure, which is likely due to accumulation of amorphous carbon and destablization by the electron beam[15-17]. In the OFF state, a gap is clearly visible in the image, while in the ON state, the gap appears considerably smaller. This provides another indication that the current flow in the device is strictly through the graphene sheet. Since the devices are suspended and can move freely, this also suggests the possibility that there is a nanomechanical component to the switching behavior.[18-21] For example, due to slack in the graphene sheet, the two halves of the original sheet that are present after the breaking procedure may be electrostatically attracted to each other when the source drain voltage is applied, completing the circuit when they come into contact. This step is expected to be relatively fast. We estimate the characteristic frequency of a graphene cantilever with the typical geometry of our devices to be ~1 MHz. This sets an upper bound for initial switch closure of ~1 µs. In this situation it is possible that there is some "switch bounce" where the layers rebound from each other after their initial contact. This motion would be damped by the intrinsic quality factor of the graphene layers, and with a typical quality factor ~300 at room temperature,[22] the motion would be reduced to the atomic scale in $t$~1 ms. Note that this estimate neglects dissipation by the collisions themselves[23] and is therefore an upper bound, likely an extreme one.

To further investigate the switching mechanism, we studied the temperature dependence of the switching rate. Figure 4 shows the time lapse to switch into the ON state when a square pulse of height 3.0V is applied, using curves with a switching time approximately equal to the average time measured for events within 1 s at each temperature, plotted versus temperature in Fig. 5a. The switching rates are highly temperature sensitive, and show a rapid increase as the temperature is raised. A histogram of the measured switching time for a range of temperatures is plotted in Fig. 5b. As the temperature is lowered, longer intervals to switch become more common, and very short intervals less common. The switching rates were also measured at lower temperatures; no switching was observed at 4.5K and indeed becomes very rare even below temperatures as large as 280 K. The long time scale for switching under appropriate conditions of >1 s greatly exceeds the upper bound estimate above for *t* of 1 ms. In addition, the rate of nanomechanical switching would be expected to have little temperature dependence over the based on the relatively constant elastic properties of the graphene sheets over the temperature range studied. This strongly indicates that the switching rate is not limited by nanomechanical motion but rather limited by a step that involves the motion of atoms and/or the rearrangement of chemical bonds that must overcome a barrier, which is in the order of ~eV, for example by the formation of carbon chains.[4,24,25] Such atomic rearrangement, if elucidated in detail, may provide the basis for long term storage. Estimates of the ultimate lifetime will require detailed understanding of the atomic motion that underlies the switching behavior. Such understating will require new experiments such as STM studies, or observation of the switching using a transmission electron microscope.

The authors acknowledge the support by NSF CAREER DMR/0748910, NSF DMR/1106358, NSF ECCS/0926056, ONR N00014-09-1-0724, ONR/DMEA H94003-10-2-1003 and the FENA Focus Center. CNL acknowledges the support by the "Physics of Graphene" program at KITP.

Supporting Information Available: *in situ* SEM videos showing the electrical breakdown process of few layer suspended graphene devices. This material is available free of charge via the Internet at http://pubs.acs.org.

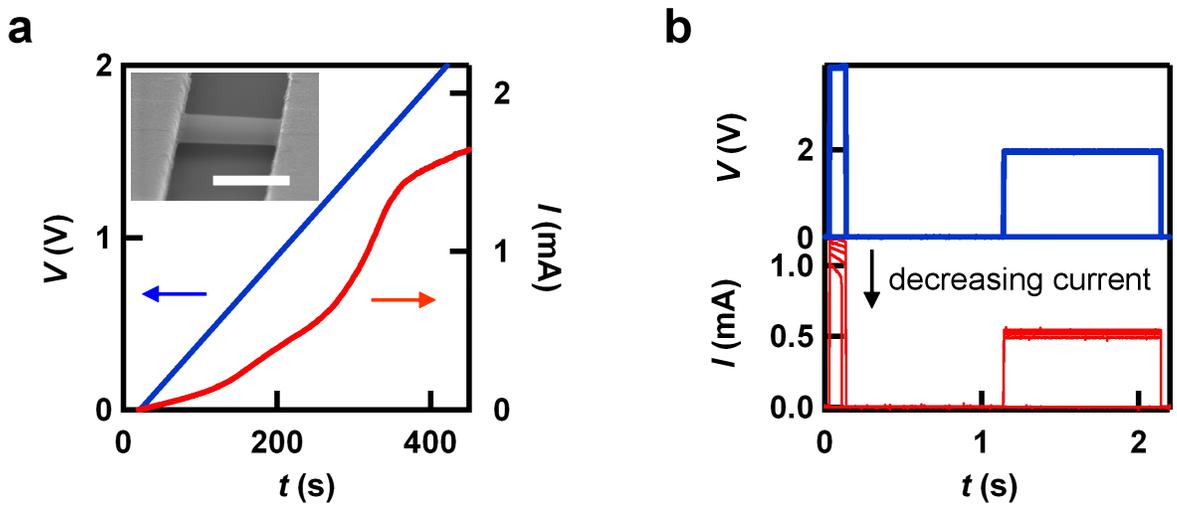

**Figure 1.** (a) The *IV*-time curve of a suspended graphene device. The blue curve indicates the bias voltage and the red curve indicates the corresponding current. The inset shows the SEM image of the device. The scale bar is 1μm. (b) (upper panel) The voltage sequence which is used to break down suspended sample. The pulse voltage is 3.9V, the test voltage after the pulse is 1.95V. (lower panel) The current response of the device during breakdown. After the last pulse, the sample is completely broken, and under the test voltage (1.95V), the corresponding current is zero.

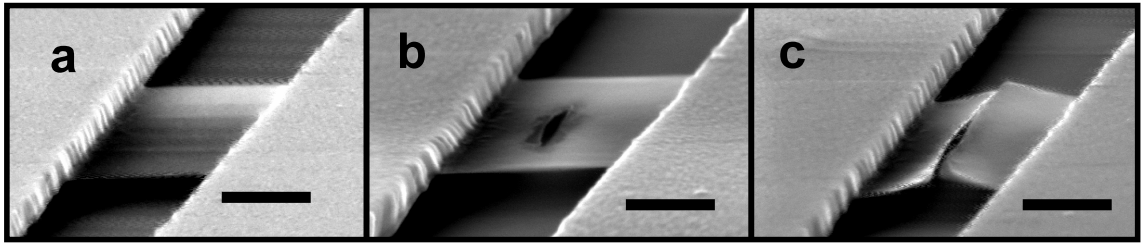

**Figure 2.** (a) A suspended sample before breakdown (b) the broken region started from the center and expanded. The image was taken after a 3V pulse was applied. (c) After increasing the amplitude of the pulse voltage, the sample was completely broken, but still free-standing. The image was taken after an 8V pulse was applied. The scale bar indicates 500nm.
.

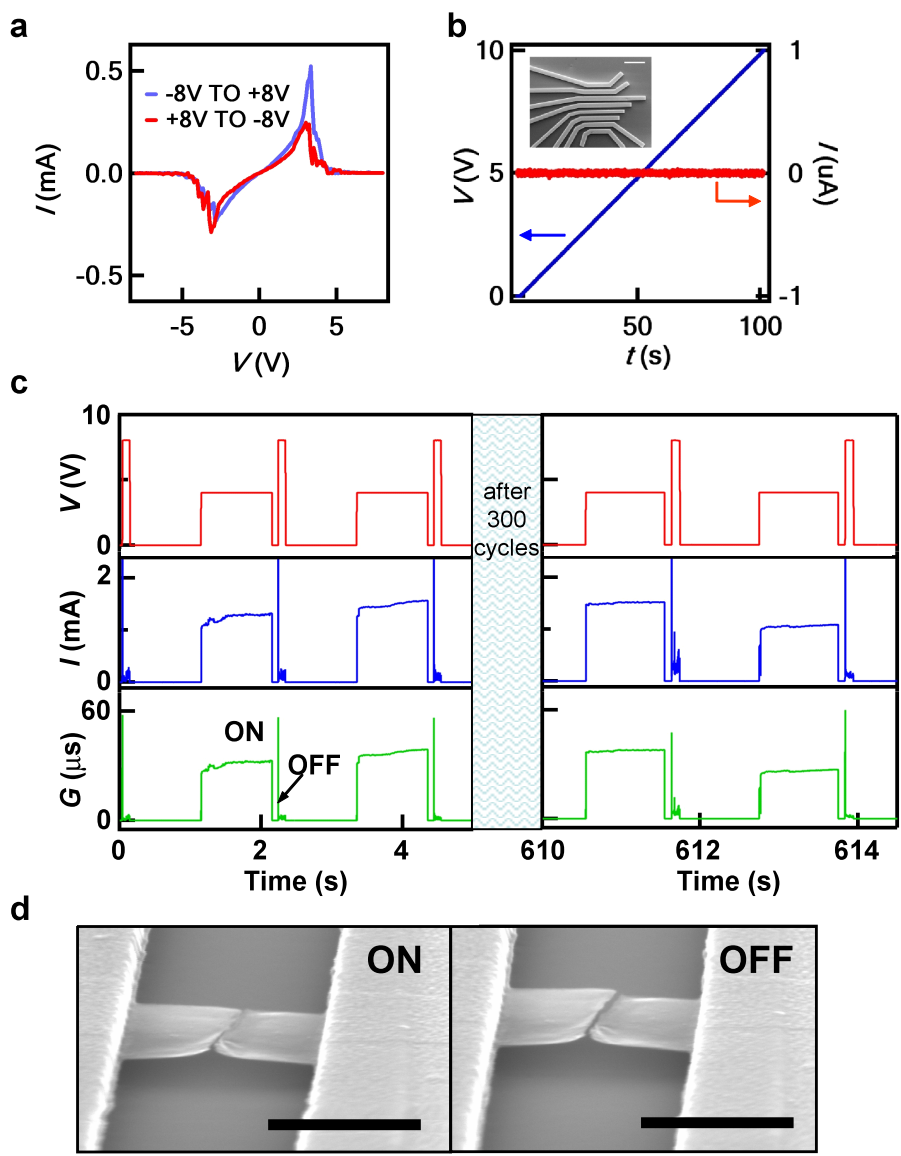

**Figure 3.** (a) The *IV* curve after breakdown. The voltage ramp from -8V to +8V is shown as the blue curve and the voltage ramp back from +8V to -8V is shown as the red curve. (b) The *IV*-time curve of a pure $SiO_2$ device which has the same geometry and the etching procedure as the device shown in Fig.1a, except no graphene sample in it. The blue curve indicates the voltage and the red curve indicates the current. We can see: during the voltage ramping up, the current is still almost zero. The inset shows the SEM image of the control device. The scale bar is 10μm. (c) The repeatability of the switching behavior. From top to bottom, the panels display the applied voltage *V*, the current response *I* and the device conductance $G=I/V$ as a function of time *t*, respectively. Left panel: initial behavior. Right panel: after 300 switching cycles. (d) The SEM images of ON and OFF state for a different device. The scale bar is 1μm.

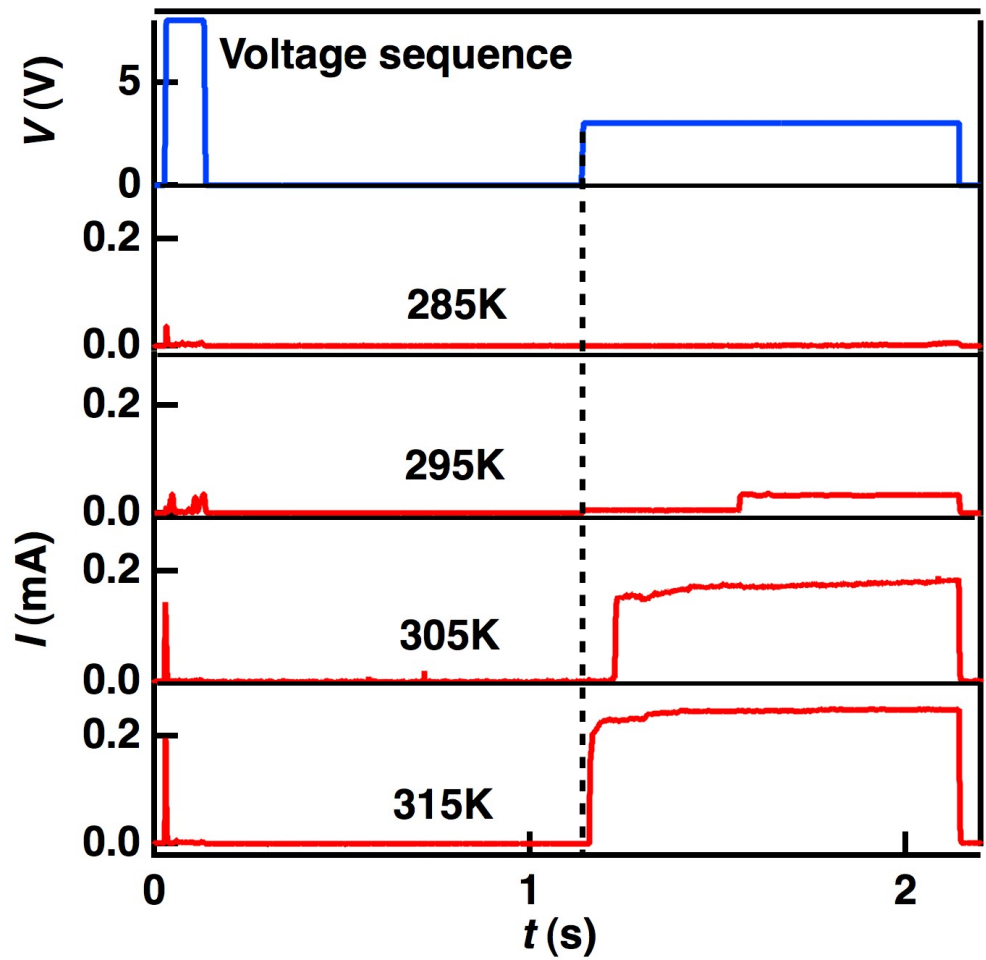

**Figure 4.** Typical current response versus time in a single switching cycle at different temperatures.

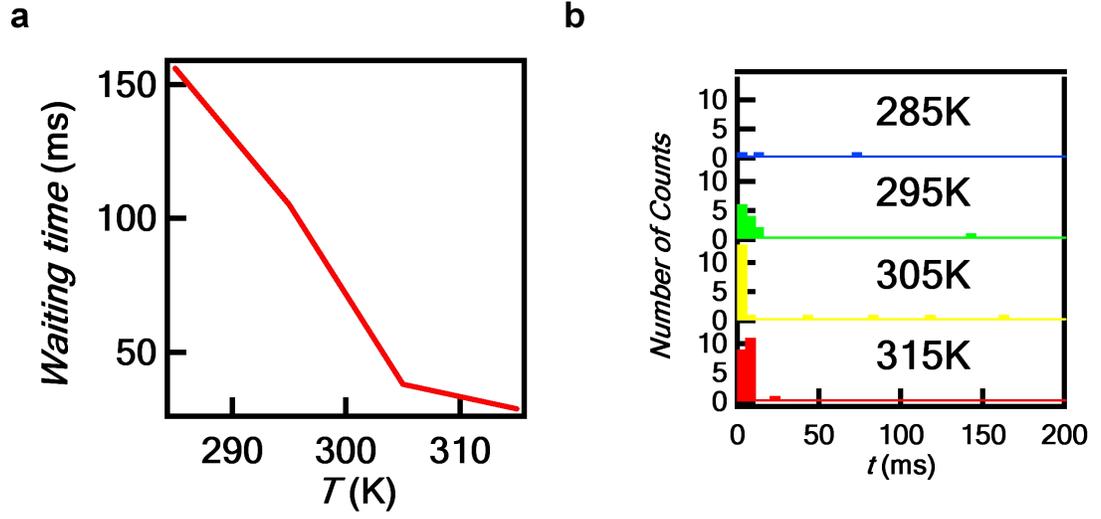

**Figure 5.** (a) Average waiting time *vs.* temperature.
(b) Statistical analysis of the waiting time.